\documentstyle[prb,aps,epsf]{revtex}
\newcommand{\be}{\begin{equation}}
\newcommand{\ee}{\end{equation}}
\newcommand{\bea}{\begin{eqnarray}}
\newcommand{\eea}{\end{eqnarray}}

\begin {document}
\title{
Breakdown of the Kondo Effect in  
Critical Antiferromagnets 
}
\author{Davide Controzzi  
}

\vspace{0.5cm}

\address{
Department of Physics, Theoretical Physics, University of Oxford,\\
1 Keble Road, Oxford OX1 3NP, UK
}

\vspace{3cm}

\address{\mbox{ }}
\address{\parbox{14cm}{\rm \mbox{ }\mbox{ }
The breakdown of the Kondo effect may be the origin of the anomalous
properties of the heavy-fermion compounds at low temperatures.  We study the 
dynamics of one
impurity embedded in an antiferromagnetic host
at the quantum critical point and show that
the impurity
is not screened and develops a
power law correlation function. This suggests that the breakdown of the
Kondo effect can simply be a consequence of the system's proximity to the quantum
critical
point.
} 
}
\address{\mbox{ }}
\address{\parbox{14cm}{\rm PACS numbers: 71.10.Hf, 71.27.+a,75.20.Hr }}
\maketitle

\makeatletter
\global\@specialpagefalse
\makeatother

A number of  heavy-fermion systems present very peculiar non--Fermi-liquid
properties which have been suggested to be a consequence of the proximity of
the antiferromagnetic  quantum critical point (QCP)\cite{andtsv}.
The weak-coupling approach to
quantum phase transitions\cite{wc},
which  describes many ferromagnetic compounds fairly well\cite{fm},
fails in
describing the temperature exponents of various thermodynamic and
transport properties of these materials\cite{fail}.
Recent
neutron scattering experiments in CeCu$_{6-x}$Au$_x$ have suggested that
the failure of this approach can be related to the
unusual energy dependence of the spin susceptibility\cite{colexp}.
The measured $\omega/T$
scaling of the spin susceptibility:
\begin{equation}
\chi (\omega,T)=\frac{1}{T^\alpha}f(\frac{\omega}{T})
\label{susc}
\end{equation}
was fitted with:
\begin{equation}
\chi^{-1}=c^{-1}[f({\bf q})+(-i\omega+aT)^\alpha]
\end{equation}
where $\alpha\simeq 0.8$, while in the weak coupling approach $\alpha=1$.
This result suggests that the anomalous non--Fermi-liquid behavior in these
compounds may  have a local origin. One possible interpretation is that
the screening that leads to the Kondo effect is incomplete as a
consequence of the proximity to the QCP \cite{colexp}.

A first analysis of the  Kondo effect in systems close to a QCP  was performed
more than twenty years ago  by Larkin and Mel'nikov\cite{lm}. They showed 
that the Kondo screening is not effective in itinerant electron
ferromagnets close to the QCP. At the present there is no theory of the
Kondo effect at the antiferromagnetic QCP.
Some authors have suggested that disorder is an essential ingredient for
the breakdown of the Kondo effect, and that it is therefore the origin of
the anomalous properties of
heavy-fermion materials. In the Kondo-disorder model
suggested by Miranda, Dobrosavljevi\'c, and Kotliar
\cite{mir}, a  randomness in the distribution of the 
hybridization matrices leads to a broad distribution of  the Kondo
temperatures with a tail that goes to zero.
Some spins then remain unscreened at all temperatures and they can be
the origin of anomalies
in the thermodynamics and transport properties. 
A different approach is based on the metallic spin glass model introduced by
Sengupta and Georges \cite{seng} and  Sachdev, Read and Oppermann
\cite{sach}.  
In these studies it was shown that, at the critical point, the spin
fluctuation spectrum is non--Fermi-liquid-like, and the impurity 
spins develop  a power
law correlation function in time.
Nevertheless the r{\^o}le of disorder in heavy-fermion materials and, in general,
of the  non-Fermi liquid properties of metals is still a hotly
debated question. It is therefore important to understand whether these results
are specific to the QCP under consideration, or if they survive also in clean
systems.

Another possible origin of the anomalous local susceptibility is the
formation of a pseudo-gap in the single particle spectrum \cite{ing}. In
quantum critical systems that show superconductivity the pseudo-gap can
be generated by superconducting fluctuations but, in general, the origin
of the pseudo-gap in critical itinerant antiferromagnets is not clear at
the moment and has been suggested mainly on the basis of studies employing
Dynamical Mean Field Theory \cite{dmft}.

One way to represent
a system when the Kondo effect breaks down  
 is to consider a spin living in a cavity in the presence of
an effective Gaussian Weiss field  
$h$ with correlation function \cite{sen,coleman}:
\begin{equation}
\left\langle Th_a(\tau)h_b(0) \right\rangle \sim
\frac{\delta_{a,b}}{\tau^{2-\epsilon}}.
\end{equation}
Integrating over the Gaussian field $h$ we get the following contribution to the effective action of the local spin:
\begin{equation}
I = g \int d\tau  \frac{S^a(\tau)S^a(\tau')}{|\tau - \tau'|^{(2 - \epsilon)}},
\label{sengact}
\end{equation}
where $g$ is the coupling constant. In itinerant antiferromagnets 
like CeCu$_{6-x}$Au$_x$ this action has to be 
combined with the usual Kondo screening. 
Such a model was analyzed by Sengupta \cite{sen}, who used an
$\epsilon$-expansion and found that the model with Kondo screening
possesses an unstable non-Fermi-liquid fixed point. In the leading
order in $\epsilon$ the exponent of the spin-spin correlation 
function  at the fixed  point coincides with the exponent for the model without Kondo screening. For the latter case the results of Sengupta coincide with a na{\"\i}ve scaling analysis which gives $[S]=\tau^{-\epsilon/2}$ and 
 $\left\langle S(\tau)S(0) \right\rangle \sim\tau^{-\epsilon}$ for the spin-spin
correlation function. In frequency 
space we have $\chi^{-1}_0(\omega)\sim (-i\omega)^{1-\epsilon}$.
This approach naturally leads to $\omega/T$ scaling of the spin susceptibility
of the form (\ref{susc}) with 
\bea
\alpha=1-\epsilon. \label{alpha}
\eea

In this paper  we discuss  the behavior of a single impurity in an
antiferromagnetic host close to the QCP. Our aim is to obtain Sengupta's
phenomenological theory from a microscopic approach to a critical
antiferromagnet. This would show that the Kondo effect in the system is
not complete. 
We first analyze   a model of itinerant electron antiferromagnets (IEAF), which
is a possible model for heavy-fermion systems, and then briefly consider also a
model for an insulator (IAF).
The
results are very similar to those obtained for metallic spin glasses and
suggest that proximity to the QCP is enough to suppress Kondo
screening.
We use a path integral approach and 
represent the bulk as a bosonic field ${\bf n(x},\tau)$ 
interacting with the impurity
via the following action:
\begin{equation}
S=S^{bulk}+J \int d\tau \: {\bf n}(0,\tau) \cdot {\bf S}(\tau),
\end{equation}
where
\begin{eqnarray}
\label{b1}
S^{bulk}=\int d\omega \, d^Dk \: [\frac{1}{2}{\bf n(k},\omega) D^{-1}_0({\bf
k},\omega){\bf n(-k},-\omega)]+ \\ \nonumber g\int d\tau \, d^D x 
\:({\bf n(x},\tau)^2)^2.
\end{eqnarray}
The form of the bare propagator, $D_0({\bf k},\omega)$, depends on the
host material. 

In
the case  of the $D$-dimensional IEAF the spin-spin correlation function of the host  
can be obtained in the random phase approximation and
is \cite{wc}:
\begin{equation}
D_0({\bf k},\omega)^{-1}=|\omega|+k^2 +\delta,
\label{prop}
\end{equation}
where $\delta$ vanishes at  the critical point. 
In relation to the bulk properties 
the ${\bf n}$-field can be treated as 
Gaussian if the effective dimensionality of the bulk 
theory exceeds the upper critical dimension. This condition is 
satisfied for the IEAF  QCP, for $D>2$ (which is the case that we shall discuss
in this paper), while for the
IAF the quartic term renormalizes the bulk propagator giving rise to an
anomalous power.

Treating the bulk field as Gaussian and integrating it out 
we obtain an effective theory of
the form (\ref{sengact}), with:
\begin{equation}
\epsilon=2-D/2, \quad \alpha = D/2 - 1. \label{dimen} 
\end{equation}
In order to check whether the interaction between the modes of the 
${\bf n}$-field, irrelevant for the bulk properties, 
remains irrelevant for the 
impurity, we have calculated the quartic term in the impurity effective action:
\begin{eqnarray}
\label{sff}
S_{\rm eff}[{\bf S}]=\frac{1}{2J}\int d\tau \: {\bf S}(\tau)D(0,\tau-\tau'){\bf
S}(\tau')+  \\ \nonumber 
g\int d\tau_1 d\tau_2 d\tau_3 d\tau_4 {\bf S}(\tau_1) \: {\bf
S}(\tau_2){\bf S}(\tau_3){\bf S}(\tau_4)
\Gamma(\tau_1,\tau_2,\tau_3,\tau_4)
\end{eqnarray}
where 
\begin{eqnarray}
\Gamma(\tau_1,\tau_2,\tau_3,\tau_4)
& = &  
\int  d\tau d\tau' d^Dx d^Dx' D({\bf x},\tau_1-\tau) \\ \nonumber
& \ & \quad D({\bf x},\tau_2-\tau) 
g({\bf x-x'},\tau-\tau')D({\bf x'},\tau-\tau_3)D({\bf
x'},\tau-\tau_4);
\end{eqnarray}
$g({\bf x},\tau)$ is the renormalized bulk vertex  and
$D({\bf x},\tau)$ the renormalized bulk propagator. As mentioned above, for 
this  model
the propagator is not renormalized and the same is true for the 
interaction, so   $\Gamma$ scales like
$\Gamma \sim \tau^{1-3/2D}$ and  the spin operator  like $S\sim
\tau^{D/4-1}$. Then  the four-spin interaction scales like
$\tau^{1-D/2}$, and  is irrelevant for $D>2$, which is the case of
interest.
 
For CeCu$_{6-x}$Au$_x$ the spin fluctuation spectrum is anisotropic in space 
\cite{anisotr} and the coefficient of the quadratic term is very small
in one direction \cite{colexp}. In the framework of this model such
anisotropy can only be obtained introducing it at a bare level. We
therefore  consider also:
\be
D_0^{-1}({\bf k},\omega)=|\omega|+Ak_\perp^2+Bk_\parallel^4+\delta
\ee
In this case we  obtain the same results with an effective dimension
given by:
\be 
D=D_{\rm eff}=5/2.
\ee
This is related to the fact that, in the action, the soft direction
provides less phase space and counts as `half a dimension'.
We then  find from (\ref{dimen}) 
that $\alpha=0.5$ in the isotropic (3D) case, and $\alpha=0.25$ in the anisotropic
one.
This simple analysis shows that  
in critical  itinerant electron antiferromagnets, described by the
model (\ref{b1}) with the bare propagator (\ref{prop}), the screening effect is
incomplete 
and the impurity develops a power law correlation function. The effect however 
is too strong to describe the observed behavior of CeCu$_{6-x}$Au$_x$. 

It is interesting to compare these results  with the behavior of an 
insulator. The QCP in such a system is described by  the (D + 1)-dimensional 
O(3) non-linear sigma model with a special value of the coupling constant. 
At the QCP the propagator has  the form:
\begin{equation}
D({\bf k},\omega)^{-1}=(\omega^2+k^2)^{1 - \eta/2}
\end{equation}
Thus in D dimensions we have:
\begin{equation}
\alpha = D - 2 +  \eta.
\end{equation}
In $D=3$, $\eta =0$ which gives $\alpha = 1$ 
(modulo logarithmic corrections). For $D = 2$ 
the index $\eta$ can be obtained by taking into account that the (2+1)-dimensional QCP 
is in the same universality class as the ferromagnetic phase 
transition point in a three-dimensional Heisenberg ferromagnet. 
For the latter system the exponent $\eta$ is known to be 0.031
\cite{itz}.
Thus in D=2 we have:
\begin{equation}
\alpha = \eta.
\end{equation}
It is reasonable to expect that for $D = 2.5$ the index $\eta$ is also small, so we get the estimate $\alpha \approx 0.5$.

In summary, we have studied the the behavior of one impurity in two
magnetic systems 
close to the antiferromagnetic quantum critical
point.
As a consequence of the proximity of the QCP, the Kondo effect
doesn't
occur, and the impurity develops a power law correlation function in
time. The emerging picture is very similar to the one proposed to
describe the non--Fermi-liquid behavior in heavy-fermion materials and to that
found in 
some works on metallic spin glasses. It suggests that a system,
approaching the QCP, undergoes a phase transition from a Kondo phase,
with Fermi liquid properties, to a non--Fermi-liquid phase,
dominated by field fluctuations.
This suggests that incomplete Kondo
screening and power law dependence of the spin-spin correlation function
are not necessarily related to disorder,
but can be simply consequences of proximity to the AFM critical point.

It is a pleasure to thank Alexei M. Tsvelik for many interesting
discussions on the subject and for comments on the draft, and
C.Hooley for reading the manuscript.


\begin{references}
\bibitem{andtsv}B.Andraka and A.M.Tsvelik, \prl {\bf 67}, 2886 (1991).
\bibitem{wc}J.A.Hertz,\prb {\bf 14}, 1165 (1976); A.J.Millis, \prb {\bf
48}, 7183 (1993).
\bibitem{colexp}Schr\"{o}der  et alt., \prl {\bf 80}, 5623.
\bibitem{fm}S.R.Julian et alt.,J.Phys:Cond.Matt.,{\bf 8}, 9675;
C.Pfleiderer et alt., \prb {\bf 55} 8330.
\bibitem{fail}See for instance H.von L\"{o}hneysen, J.Phys:Cond.Matt.,{\bf
8},9689.
\bibitem{lm}A.I.Larkin and V.I.Melnikov, JEPT, {\bf 34}, 656 (1972).
\bibitem{mir}E.Miranda, V.Dobrosavljevi\'{c}, G.Kotliar 
\prl {\bf 78}, 290 (1997).
\bibitem{seng}A.M.Sengupta and A.Georges, \prb {\bf 52}, 10295 (1995).
\bibitem{sach}S.Sachdev, N.Read and R.Oppermann \prb {\bf 52}, 10286
(1995).
\bibitem{sen}A.Sengupta, cond-mat 9707316.
\bibitem{coleman}P.Coleman, Proceeding to SCES98, Physica B {\bf 353},
258 (1999). 
\bibitem{ing}K.Ingersent and Q.Si, cond-mat/9810226; Q.Si, J.L.
Smith and K.Ingersent, Int.J.Mod.Phys.B {\bf 13}, 2331 (1999).
\bibitem{dmft}A.Georges, G.Kotliar, W.Krauth and J.Rozenberg,
Rev. Mod.Phys {\bf 68}, 13 (1996).
\bibitem{anisotr} A.Rosch, A.Shr\"{o}der, O.Stockert and 
H.von L\"{o}hneysen, \prl
{\bf 79}, 159 (1997).
\bibitem{itz}C.Itzykson and J-M.Drouffe, Statistical field theory,
Cambridge University Press (1989).
\end{references}
\end{document}